\RequirePackage{fix-cm}
\documentclass{svjour3}                     % onecolumn (standard format)
\smartqed  % flush right qed marks, e.g. at end of proof
\usepackage{graphicx}
\usepackage{mathptmx}      % use Times fonts if available on your TeX system
\usepackage{latexsym}
\journalname{Experimental Astronomy}
\begin{document}

\title{A Systematic Analysis of the {\it XMM-Newton} Background:\\II. Properties of the in-Field-Of-View Excess Component
%Characterizing the {\it XMM-Newton}/EPIC Particle Background%\thanks{Grants or other notes}
}
%\subtitle{II. Properties of the in-Field-Of-View Excess Component}%Do you have a subtitle?\\ If so, write it here}

\titlerunning{Properties of the in-Field-Of-View Excess Component}        % if too long for running head

\author{David Salvetti \and
	Martino Marelli$^{1}$ \and
	Fabio Gastaldello$^{1}$ \and
	Simona Ghizzardi$^{1}$ \and
	Silvano Molendi$^{1}$ \and
	Andrea De Luca$^{1,3,4}$ \and	
	Alberto Moretti$^{2}$ \and	
	Mariachiara Rossetti$^{1}$ \and
	Andrea Tiengo$^{1,3,4}$
}


\institute{David Salvetti \at
              $^1$INAF-Istituto di Astrofisica Spaziale e Fisica Cosmica, via E. Bassini 15, I-20133 Milano, Italy\\
              Tel.: +39-02-23699617\\
              \email{salvetti@iasf-milano.inaf.it}             \and
	$^{2}$ INAF-Osservatorio Astronomico di Brera, via Brera 28, I-20121 Milano, Italy \\
$^{3}$ Scuola Universitaria Superiore IUSS Pavia, piazza della Vittoria 15, I-27100 Pavia, Italy  \\
$^{4}$ Istituto Nazionale di Fisica Nucleare, Sezione di Pavia, via A. Bassi 6, I-27100 Pavia, Italy
}

\date{Received: date / Accepted: date}

\maketitle

\begin{abstract}
We present an accurate characterization of the particle background behaviour on {\it XMM-Newton} based on the entire EPIC archive. This corresponds to the largest EPIC data set ever examined. Our results have been obtained thanks to the collaboration between the FP7 European program EXTraS and the ESA R\&D ATHENA activity AREMBES. We used as a diagnostic an improved version of the diagnostic which compares the data collected in unexposed region of the detector with the region of the field of view in the EPIC-MOS. We will show that the in Field-of-View excess background is made up of two different components, one associated to flares produced by soft protons and the other one to a low-intensity background. Its origin needs to be further investigated.
%Insert your abstract here. Include keywords, PACS and mathematical
%subject classification numbers as needed.
\end{abstract}

\section{Introduction}\label{intro}

{\it XMM-Newton} is an X-ray observatory \cite{jan01} launched on December 10th, 1999. Its main instrument is the European Photon Imaging Camera (EPIC), consisting of two MOS detectors \cite{tur01} and a pn camera \cite{str01} which operate in the 0.2--12 keV energy range. The EPIC background can be separated into particle, photon and electronic noise components (see \cite{car07} for a more accurate description).

The aim of this work is to accurately describe and characterize the instrumental particle-induced background concentrating on the focused one, which is generated by the interaction of low-energy particles (E$<$100 keV) with the detectors. This background component is characterized by an extreme time variability, ranging from $\sim$100 s to several hours, where the peak count rate can be more than three orders of magnitude higher than the quiescent one. Since the involved particles are protons of low energy, such episodes are known as soft proton (SP) flares. To date, our comprehension of these processes on board {\it XMM-Newton} is still incomplete.

There has been considerable work over the years on this topic, e.g. \cite{del04} measured the contamination given from SP to characterize the cosmic X-ray background; \cite{car07} studied the various components to EPIC background; \cite{kun08} characterized the spectral and spatial response of the EPIC-MOS detector to SP background; \cite{lec08} characterized all background components, included SP one, to study the radial temperature profiles for galaxy clusters; \cite{fra14} analysed EPIC background in order to investigate the potential solar axion signature in a large EPIC dataset spanning over a 12-year period. 

Thanks to the collaboration between AREMBES\footnote{\texttt{http://space-env.esa.int/index.php/news-reader/items/AREMBES.html}} and EXTraS\footnote{\texttt{http://www.extras-fp7.eu}} \cite{del15} projects, we have performed a systematic analysis of the entire EPIC archive for the first time. Our work is based on 13 years of {\it XMM-Newton} observations, from 2000 to 2012. This corresponds to the largest EPIC data set ever analysed. This allows studying and characterizing meticulously the behaviour of the excess particle background in the detector area exposed to sky photons through on spectral and timing analysis.

We define two detector areas, the in-Field-Of-View ({\it inFOV}) one, exposed to focused X-ray photons, and the out-Field-Of-View ({\it outFOV}) one, not exposed to sky photons \cite{mar}. To estimate the {\it inFOV} excess particle background in the EPIC instrument, we have mostly made use of the ``{\it inFOV} subtracted by {\it outFOV}'' diagnostic. This approach is slightly different with respect to what done by previous authors who used the ``{\it inFOV} over {\it outFOV}'' diagnostic. The latter method is thorough if the purpose is to quantify the contribution of the {\it inFOV} excess particle background to EPIC background or to analyse its spectral behaviour, but only with the former diagnostic is it possible to describe in detail its characteristics.

Due to the lack of a proper {\it outFOV} for the pn detector, this approach can only be performed on MOS cameras. We have chosen to study only the background on MOS2 because the MOS1 data set is not complete due to the loss of a CCD in 2005 after the impact with a micrometeoroid. Statistical quality of the data is unprecedented. To fully exploit this we have performed a scrupulous analysis of systematic errors, which are often the source of the dominating uncertainties in our work. 

This work is part of the project within AREMBES aimed at studying the behaviour of particle-induced background on {\it XMM-Newton}. Several parts of this project are reported in these proceedings \cite{mar,ghi,gas}. In \cite{mar} primary definition and filters on our data set are given, while \cite{ghi} employ information provided by this work to study in detail the behaviour of the {\it inFOV} excess particle background as a function of the position in the terrestrial magnetosphere. In the end, \cite{gas} focus on the study and characterization of the behaviour of the unfocused and focused particle-incuded background.

The proceeding is organized as follows: in Sect.~\ref{sec:2} we describe in detail the pipeline to extract the {\it inFOV} count rate subtracted by {\it outFOV} one from the largest EPIC data set. In Sect.~\ref{sec:3} we present our results, and we discuss the nature of a new component in the {\it inFOV} excess particle background in Sect.~\ref{sec:4}.

\section{The pipeline to extract the {\it inFOV} excess particle background}\label{sec:2}
%Text with citations \cite{RefB} and \cite{RefJ}.

We have developed an ad-hoc pipeline to extract the {\it inFOV} excess particle background from EPIC data. The initial data set consists of cleaned light curves for the {\it inFOV} and the {\it outFOV} with time bin of 500 sec extracted from a sub-sample of all 7427 public exposures performed from February 3rd, 2002, to December 8th, 2012 (see \cite{mar} for a detailed description of the analysis).

As previously described, the best way to study the behaviour of the {\it inFOV} excess particle background is to focus on the {\it inFOV} employing the {\it outFOV} region as a calibrator to minimize any contamination. For this reason we produce an {\it outFOV}-subtracted {\it inFOV} light curve where in each 500 sec time bin the count rate is the difference between {\it inFOV} and {\it outFOV} count rate while the 1$\sigma$ standard deviation is calculated using the standard error propagation rule. Since {\it outFOV} light curves have poor statistic in a time bin of 500 sec, we calculate a running mean, a running standard deviation and a running fractional exposure using the 2 nearest time bins around each data point. These light curves are characterized by time bins of 2500 sec and step of 500 sec. For the first and last 2 time bins we force the count rate and its standard deviation to --1, while the fractional exposure to 0.

We merge all the generated {\it inFOV}-{\it outFOV} light curves to obtain a final light curve for the sky fields. The most important product of our pipeline is a file containing per each 500 sec time bin the most important information to study and characterize the {\it inFOV} excess particle background.

We exclude from the analysis time bins where the counts statistic is too poor in the {\it inFOV} or in the {\it outFOV} to apply the Gaussian statistic. The number of observed counts in each time bin depends strongly on two parameters, the fraction of the bin that is exposed  ({\it F}) and the fraction of the area that is exposed ({\it R}). Increasing {\it F} the photon counts increase, while increasing {\it R} they decrease. A strong estimator to filter time bins with poor statistic may be extracted from the distribution of the ratio between {\it F} and {\it R} for {\it inFOV} and {\it outFOV}. Analysing such distribution, we decide to include in our data set only time bins with a ratio greater than 0.49 for {\it inFOV} and than 0.29 for {\it outFOV}.

\section{Results}\label{sec:3}

\begin{figure*}
\centering
  \includegraphics[width=0.85\textwidth]{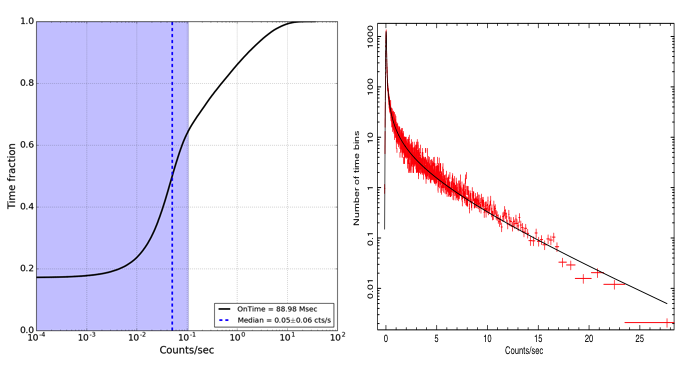}
\caption{(Left) {\it inFOV}-{\it outFOV} count rate cumulative distribution function. Blue dotted vertical line shows the median, while the light blue span the median absolute deviation. (Right) {\it inFOV}-{\it outFOV} count rate differential distribution function. Black line represents the best-fit model as described in the text}
\label{fig:1}
\end{figure*}

The final filtered data set is characterized by an exposure time of $\sim$88.98 Msec. To date, this is the largest data set employed to study the {\it inFOV} excess particle background detected by {\it XMM-Newton} EPIC, which allows us to characterize it in a detail never achieved before. From such data set we construct the count rate cumulative distribution function (CDF) and the differential distribution function (DDF) (see Figure~\ref{fig:1}). The former shows the fraction of time with respect to the filtered exposure time (here named ``OnTime'') spent below a given count rate, while the latter shows the number of time bins where the count rate is within a given count rate bin. We analyse such distributions to characterize in unprecedented detail the {\it inFOV} excess particle background. 

\subsection{{\it inFOV}-{\it inFOV} count rate distribution}\label{sec:3.1}

Count rate CDF and DDF clearly show that the {\it inFOV} excess particle background is composed by two main components, a ``{\it low intensity}'' one, characterized by a low count rate distributed following a Gaussian shape, and a ``{\it flaring}'' one, characterized by an higher count rate distributed following a more complex shape, similar to a power law. 
Analysing the CDF we can extract the fraction of observing time with flares. The flaring component becomes dominant in the distribution for a count rate larger than $\sim$0.1 cts/s. The fraction of time when the {\it inFOV} excess particle background is characterized by a count rate larger than such value is $\sim$35\% of ``OnTime'' (31.15 Msec). For the remaining $\sim$65\% of time (57.83 Msec) the {\it inFOV} excess particle background is dominated by the low intensity component.

\subsubsection{Empirical characterization of the {\it inFOV} excess particle background}\label{sec:3.2}

Analysing the DDF we can study in detail the shape of the {\it inFOV} excess particle background components. The distribution is characterized by a Gaussian component in addition to a more complex one. We model the latter component with an empirical model defined as a modified Lorentzian distribution as follows:
\begin{equation}
	F(x)=\frac{LN\cdot x^{\Gamma_1}}{1+\left|\frac{2(x-LC)}{LW}\right|^{\Gamma_2}}\cdot e^{x/X_0}
\end{equation}
where {\it LN} is the normalization and {\it LC} the center of the Lorentzian, {\it LW} the full width at half maximum (FWHM), {\it $\Gamma_1$} the slope of power-law component, {\it $\Gamma_2$} the slope of the denominator component and {\it $X_0$} the exponential cut-off component.

We find that the Gaussian component is characterized by a mean value significantly different from zero (0.0174$\pm$0.0001 cts/s) and a standard deviation equal to 0.0328$\pm$0.0001 cts/s.  The width of the Gaussian is related to the subtraction process and is associated with the statistical fluctuation at low count rate

The values of parameters related to the modified Lorentzian component are below:
LC=0.079$\pm$0.001; LW=0.110$\pm$0.001; $X_0$=5.37$\pm$0.06; $\Gamma_1$=0.47$\pm$0.03; $\Gamma_2$=1.34$\pm$0.03. All quoted errors on the empirical model parameters are at 1$\sigma$ confidence level for a single interesting parameter.

\subsection{{\it inFOV}-{\it outFOV} distribution as a function of filter}\label{sec:3.3} 

\begin{figure*}
\centering
  \includegraphics[width=0.55\textwidth]{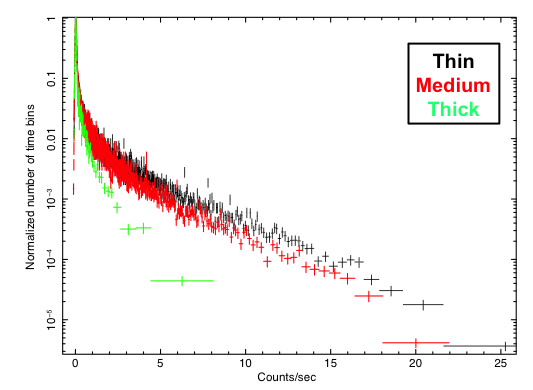}
\caption{Normalized {\it inFOV}-{\it outFOV} count rate differential distribution function for observations with Thin (black), Medium (red) and Thick (green) filter.}
\label{fig:3}
\end{figure*}

Starting from our filtered data set, we produce the count rate differential distribution for each EPIC optical filter (see Figure~\ref{fig:3}). Focusing on the high count rate region, which is dominated by flaring background component, we observe that the DDF is quite similar for the Thin and Medium filter, while it is very different for the Thick filter. Flaring background has a different response as a function of the EPIC optical filter. This behaviour is one of the main indications we have that the flaring component is composed of soft protons. Indeed, soft protons are more affected by the Thick than the Medium or Thin filters (see \cite{tie07}).

Conversely, if we focus on the low count rate region, which is dominated by the low-intensity background component, we observe that the distributions do not seem so different. This is at variance with what we would expect if the low intensity component were indeed associated to soft protons. Its origin needs to be further investigated.

\section{Investigating the nature of the low intensity background component}\label{sec:4}

\subsection{Evaluating systematics effects}\label{sec:4.1}

Having identified a low intensity component in the {\it inFOV} excess particle background distribution, we have to check if this may be explained by a systematic effect. We estimate systematics effects related to the subtraction procedure of {\it outFOV} data to {\it inFOV} through the study of the observations with the filter wheel in closed position. In this configuration, an aluminium window prevents X-ray photons and low energy particles from reaching the detectors. Since the instrumental background dominates these exposures, we extract {\it inFOV}-{\it outFOV} light curves using the pipeline described in Sect.~\ref{sec:2} to investigate and calculate a possible counts excess in the {\it inFOV} region because of an inhomogeneous distribution of the internal instrumental background on the MOS2 camera.

\begin{figure*}
\centering
  \includegraphics[width=0.85\textwidth]{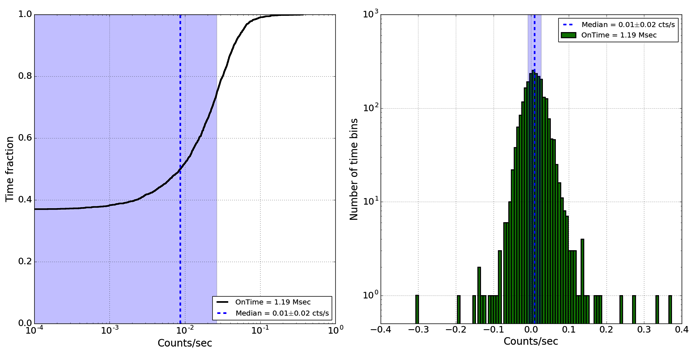}
\caption{(Left) {\it inFOV}-{\it outFOV} count rate cumulative distribution function for the closed exposures. Blue dotted vertical line shows the median, while the light blue span the median absolute deviation. (Right) {\it inFOV}-{\it outFOV} count rate differential distribution function.}
\label{fig:2}
\end{figure*}

We have retrieved 72 closed observations (corresponding to 73 exposures) from the official list on the {\it XMM-Newton} web page\footnote{\texttt{http://xmm-tools.cosmos.esa.int/external/xmm\_calibration/background/filter\_closed/mos/mos2/mos2\_FF\_2016\_v1.shtml}}. 
We apply the automatic pipeline and create the {\it inFOV}-{\it outFOV} count rate CDF and DDF (see Figure~\ref{fig:2}). Both distributions clearly show that there is no SP flaring background in closed exposures as we expect because only the Gaussian component at low count rate is detected. Fitting a simple Gaussian model to DDF distribution, we obtain that the best-fit mean value is 0.0085$\pm$0.0006, while the standard deviation is 0.0260$\pm$0.0005. This result shows an excess of counts from instrumental background in the {\it inFOV} region. We have not investigated in detail the origin of such excess but analysing the integrated image containing all the closed exposures we assert that the major contribution may be due to the process of the electronic readout, which produces an asymmetric distribution of electronic background in each CCD in the direction of the readout nodes.

If we include such a systematic effect in our analysis, we find that the importance of the low intensity component for the non-closed exposures decreases but remains significant. Indeed, the magnitude of the low intensity component is roughly twice that which can be attributed to the systematic effect in the subtraction procedure. This result confirms the presence of an unexpected background component that is characterized by a very low count rate with respect to the SP flaring background component.

Considering that the rescaled {\it outFOV} intensity is $\sim$0.02 cts/s (as we will show in the next section), a systematic error of $\sim$0.008 cts/s in the subtraction procedure corresponds to a relative systematic error of $\sim$4\%, we take this as a rough estimate of the magnitude of systematic effects that are affecting our measures.

\subsection{Unfocused high-energy particles scenario}\label{sec:4.2}

\begin{figure*}
\centering
  \includegraphics[width=0.85\textwidth]{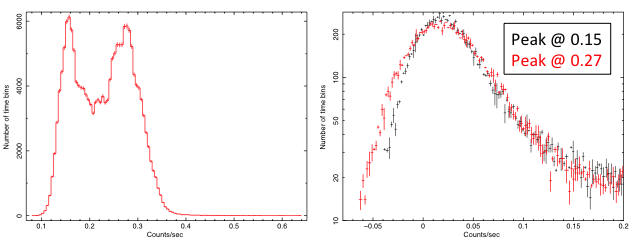}
\caption{(Left) {\it outFOV} count rate differential distribution. The bimodal distribution is associated to the solar cycle. (Right) : {\it inFOV}-{\it outFOV} count rate differential distribution function during the period of two peaks in the {\it outFOV} one. The distribution referred to the period characterized by high count rate is shown in red while the period characterized by low count rate in black. The two distributions were renormalized in order to have the same peak value.}
\label{fig:4}
\end{figure*}

The left panel of Figure~\ref{fig:4} shows the count rate DDF for the {\it outFOV}. This is characterized by a two-peaks distribution at low count rate, with the first located at $\sim$0.16 cts/s and the second at $\sim$0.28 cts/s. Such distribution is clearly associated with the modulation of unfocused particle background produced during the solar cycle \cite{gas}. The flaring background signal, characterized by high count rate, is not seen in the {\it outFOV} as expected. 

If the low intensity component of the {\it inFOV}-{\it outFOV} distribution were related to a residual of the unfocused high-energy particle (E$>$100 MeV) component we would expect its distribution, corresponding to the period of the two peaks, to change because of the different contribution of high-energy particles to the two states. We extract {\it inFOV}-{\it outFOV} DDF corresponding to count rate between 0.14 and 0.17 cts/s for the {\it outFOV} (first peak), and to 0.26 and 0.29 cts/s (second peak). The right panel of Figure~\ref{fig:4} shows the two renormalized distributions.

As we expect from the different statistics in the two peaks, the width of the distributions is different, fitting the entire distribution with our empirical model we obtain a standard deviation value of 0.0266$\pm$0.0001 cts/s for the first peak and 0.0308$\pm$0.001 cts/s for the second one. The best-fit value of two Gaussian means is 0.0129$\pm$0.0002 for the first peak and 0.0162$\pm$0.002 for the second. As discussed in Sect.~\ref{sec:4.1}, these values are compatible including the contribution of the systematic error making unlikely an unfocused high-energy particle nature for the low intensity component.

\subsection{Evolution through the mission}\label{sec:4.4}

We want to test if the {\it inFOV} excess particle background given by the newly discovered low intensity background component shows an evident evolution through the {\it XMM-Newton} mission and if such evolution is different from the evolution of flaring component.

Starting from our filtered data set, we have studied the {\it inFOV}-{\it outFOV} light curves dividing data per year. In this way we have extracted and analysed count rate CDF and DDF for 13 years of mission, from 2000 to 2012. Obviously this is a simple approach that aims at investigating the {\it inFOV} excess particle background evolution on time scale of several years. A more accurate analysis that takes into account the behaviour of the {\it inFOV} excess particle background components as a function of the position of the satellite in the terrestrial magnetosphere is describe in \cite{ghi}.

\begin{figure*}
\centering
  \includegraphics[width=0.5\textwidth]{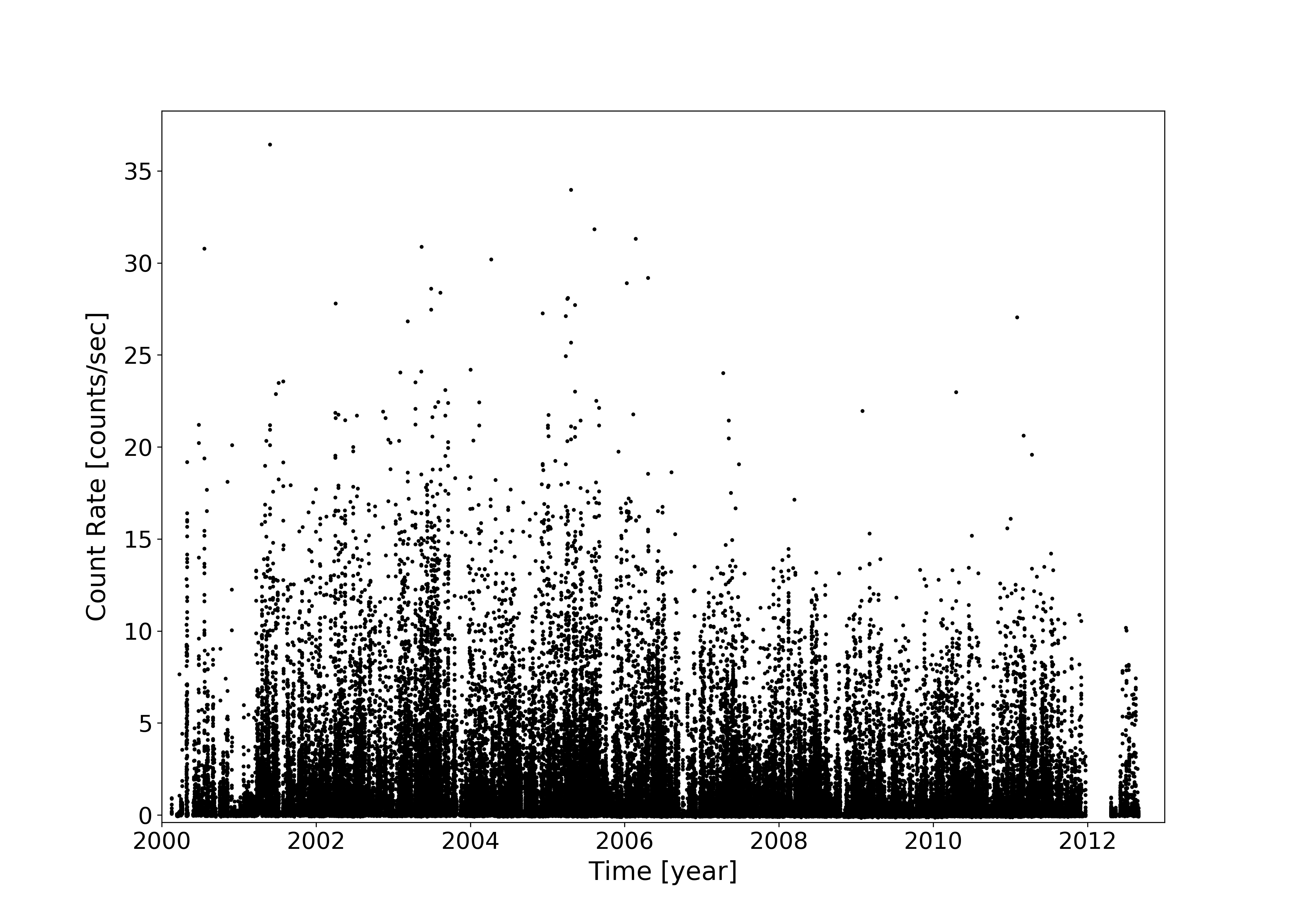}\hspace{-.5cm}
 \includegraphics[width=0.5\textwidth]{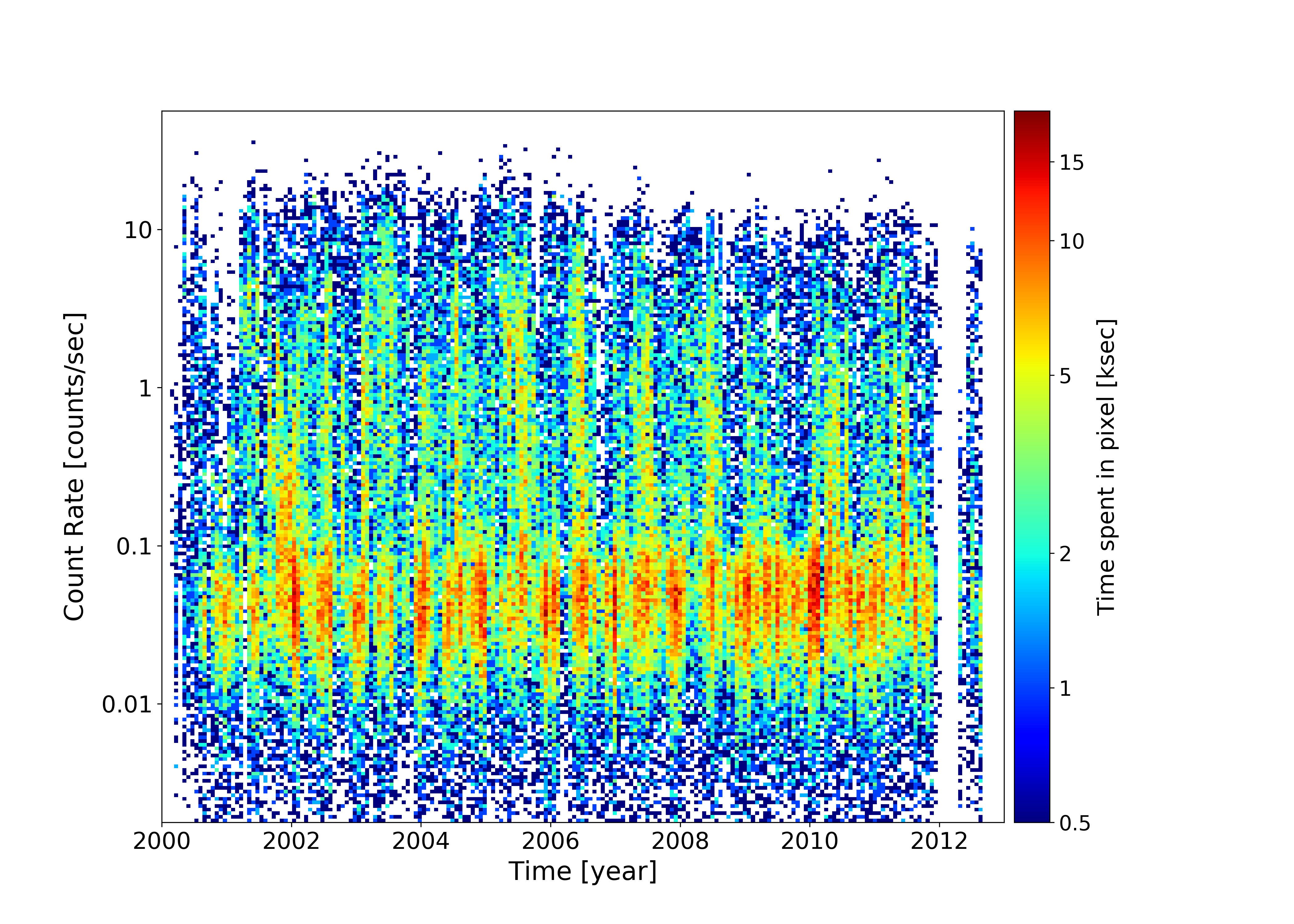}
\caption{{\it inFOV}-{\it outFOV} MOS2 light curves for 13 years of {\it XMM-Newton} mission, from 2000 to 2012. (Left) Linear scale on count rate axis shows evidence for evolution of the SP flaring background through the mission. (Right) Logarithmic scale on count rate axis focuses on the low-intensity background component (located in the denser region).}
\label{fig:5}
\end{figure*}

Figure~\ref{fig:5} shows the {\it inFOV}-{\it outFOV} light curve for 13 years of {\it XMM-Newton} mission. The plot on the left shows the time evolution of the intensity and importance of SP flaring background component, while the plot on the right focuses on the time evolution of the low-intensity background component.
We find some indication of evolution for the SP flaring component through the mission, conversely no clear variation of the low intensity component can be significantly detected.

\subsection{Spectral analysis}\label{sec:6}

\begin{figure*}
\centering
  \includegraphics[width=0.85\textwidth]{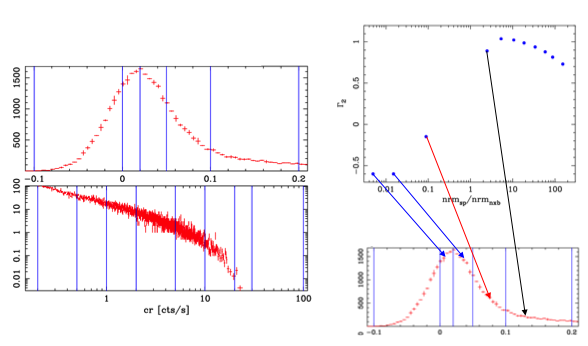}
\caption{(Left) {\it inFOV}-{\it outFOV} count rate differential distribution function for the medium filter. Blue vertical lines indicate the ranges over which {\it inFOV} and {\it outFOV} spectra were extracted. (Right) Top panel, high energy spectral slope of the {\it bkn2} component as a function of the ratio of normalizations of the {\it bkn2} and {\it bkn1} components.  Bottom panel, differential distribution of the {\it inFOV} excess background. Note how the slope of the {\it bkn2} component rapidly changes as we move from the peak region, dominated by the low intensity component (blue arrows), to the high count rate region, dominated by soft proton flares (black arrow).}
\label{fig:6}
\end{figure*}

Spectral analysis of the data can provide further insight into the nature of the low intensity component discussed in the previous subsections. We have extracted spectra from the {\it inFOV} and {\it outFOV} regions for different levels of {\it inFOV}-{\it outFOV} intensity so as to separate as much as possible contributions from the low intensity component and SP flares. For each filter we have extracted 13 spectra, the choice of limiting {\it inFOV}-{\it outFOV} values are reported in the left panel of Figure~\ref{fig:6} together with the {\it inFOV}-{\it outFOV} distribution. As we can observe in the figure, we have a sufficiently large number of spectra to follow the transition from the low intensity contaminated region to the SP flare dominated region. We have performed spectral analysis using XSPEC\footnote{\texttt{https://heasarc.gsfc.nasa.gov/xanadu/xspec/manual/XspecManual.html}} v12.9 software.

At variance with what we have done for the lightcurve analysis, we have not subtracted the {\it outFOV} spectrum from the {\it inFOV} one, but, as in \cite{lec08}, we have worked with models. More precisely we have built a 4 component model comprising: 1) a first broken power-law component, {\it bkn1}, accounting for the high energy particle induced component observed both in the {\it inFOV} and the {\it outFOV} regions; 2) a multi-gaussian component, {\it mgau}, accounting for the many fluorescence lines observed in the the {\it inFOV} and the {\it outFOV} regions; 3) a second broken power-law component, {\it bkn2}, accounting for the excess emission observed in the {\it inFOV} region only  and finally 4) a cosmic X-ray background component, {\it cxb}, for the cosmic X-ray emission observed in the the {\it inFOV} region only. Fitting was performed simultaneously on each {\it inFOV} and {\it outFOV} spectra pair. Parameters for the {\it bkn1} component were forced to be the same for the two spectra, for the {\it mgau} component only energies were tied together while the normalizations were left to vary freely from one another to allow for variations of fluorescence lines across the detector. Spectral fits were performed for all spectra and for all filters.

Once the analysis is performed, evolution of spectral parameters can be used to characterize the behavior of the various components. Since we are interested in the {\it inFOV} contamination, we have examined the {\it bkn2} component. In the right panel of Figure~\ref{fig:6} we show the variation of the high energy spectral slope of bkn2 as a function of the ratio of the normalization of {\it bkn2} over {\it bkn1}, i.e. {\it nrmbkn2/nrmbkn1}. The first parameter describes the spectral shape of the {\it inFOV} contamination while the second is a measure of  its intensity relative to that of the high energy particle induced component. We can identify 3 different  regions: a region associated to the peak (blue arrows in the right panel of Figure~\ref{fig:6}) where the {\it inFOV} contamination is dominated by the low intensity component; a region at high count rates (black arrow in the right panel of Figure~\ref{fig:6}) dominated by the flaring soft proton component and an intermediate region (red arrow in the right panel of Figure~\ref{fig:6}) where both components contribute. As we can see from the top panel of the right panel of Figure~\ref{fig:6} in the first region the spectral slope is very flat, $\sim$ $-$0.6, in the second it is much steeper, $\sim$1 and in the intermediate region it undergoes a very rapid transition from one regime to the other. These results suggest that the low intensity and soft proton components are different in nature, this is in agreement with what has been found from the analysis of the {\it inFOV}-{\it outFOV} distribution as a function of filter, see Sect.~\ref{sec:3.3}. Preliminary {\it GEANT4} simulation results are showing that low intensity component may be produced by Compton interaction of hard X-ray photons with the telescope \cite{hal10}.

\section{Conclusions}

In this work we have described and characterized the {\it inFOV} excess particle background on EPIC MOS2 camera on board {\it XMM-Newton}. The statistical quality of data is unprecendented: we have analysed 13 years of observations, from 2000 to 2012. We have used {\it outFOV} region as a calibrator to minimize any contamination. For this reason we have produced and studied {\it outFOV}-subtracted {\it inFOV} light curves with a time bin of 500 sec. Excluding from the analysis ``bad'' exposures and time bins, our final data set is roughly 90 Msec.

Analysing the count rate cumulative distribution function of {\it inFOV}-{\it outFOV} light curves we have measured the fraction of the flaring time in {\it XMM-Newton} MOS2 is about 35\% ($\sim$30 Msec). The count rate differential distribution functions shows two component in the background, one associated to flares and the other to a low intensity component.

A comparative analysis of data collected with different filters shows that the flaring component is consistent with being produced by protons in the tens of keV range, while the low intensity one is not. A dedicated analysis shows that only about half of the low intensity component can be attributed to systematics in the subtraction process. 

A spectral analysis of our data confirms that the flaring and the low intensity components differ in nature. Intriguingly, while the evidence we now have is enough to state with some certainty, that the low intensity component is not associated to soft protons, it is still insufficient to say more about its nature. This unexpected result has significant implications in terms of our understanding of the {\it XMM-Newton}/EPIC background. Recent {\it GEANT4} simulations are showing that Compton interaction of hard X-ray photons with the telescope may be the origin of the low-intensity component. Deeper analyses are necessary to confirm such hypothesis.

%\subsubsection{subsubsection headings} Use subsubsection headings as needed.

%\begin{figure}
 % \includegraphics{example.eps}
%\end{figure}
%\begin{figure*}
%  \includegraphics[width=0.75\textwidth]{example.eps}
%\end{figure*}
%\begin{table}
%\begin{tabular}{lll}
%\hline\noalign{\smallskip}
%\end{tabular}
%\end{table}

\begin{acknowledgements}
The AHEAD project (grant agreement n. 654215) which is part of theEU-H2020 programm is acknowledged for partial support. This work is part of the AREMBES WP1 activity funded by ESA through contract No. 4000116655/16/NL/BW. Results presented here are based, in part, upon work funded through the European Union Seventh Framework Programme (FP7-SPACE-2013-1), under grant agreement n. 607452, ``Exploring the X-ray Transient and variable Sky - EXTraS''. We thank the anonymous referee for his/her very helpful comments to our manuscript. 
\end{acknowledgements}



\end{document}